\documentclass[preprint,12pt]{elsarticle}

\usepackage{amssymb}
\usepackage{amsmath}
\usepackage{booktabs}
\usepackage{xcolor}
\usepackage{tikz}
\usetikzlibrary{arrows.meta,positioning,fit,backgrounds}
\usepackage{lineno}
\usepackage[hidelinks]{hyperref}
\journal{Engineering Applications of Artificial Intelligence}

\begin{document}

\begin{frontmatter}

\title{Removing Speech, Keeping Activities: Speech Removal for Privacy-Preserving Acoustic Sensing in Assisted Living}

\author[inst1]{Pavlos Nicolaou}
\author[inst2]{Christos Efstratiou}

\affiliation[inst1]{organization={KIOS Research and Innovation Center of Excellence, University of Cyprus},
    addressline={Panepistimiou 1, Aglantzia}, 
    city={Nicosia},
    postcode={2109},
    country={Cyprus}}

\affiliation[inst2]{organization={School of Computing, University of Kent},
    city={Canterbury},
    postcode={CT2 7NZ}, 
    state={Kent},
    country={United Kingdom}}

\begin{abstract}
Acoustic sensing can monitor the daily activities of older adults without cameras or wearables, yet the possibility of capturing private conversations remains a barrier to its deployment. The engineering application addressed in this paper is privacy-preserving acoustic monitoring of daily activities in assisted living. Our contribution in artificial intelligence is a speech-removal front end: a U-Net that masks the short-time Fourier transform of ambient audio and returns a waveform in which speech is suppressed and the sounds needed for activity recognition are kept. The model is trained only on synthetic mixtures of public sounds and speech, so no speech has to be recorded in the target homes. We evaluate it with group-disjoint cross-validation on environmental and domestic sound datasets mixed with reverberant speech, on speech from unseen corpora and on real home recordings, and measure privacy with three voice activity detectors, Whisper speech recognition and a speaker-verification attack. With speech 5~dB above the background, the macro-averaged F1 score of activity recognition is 0.574 and 0.661 after removal on the two datasets, compared with 0.233 and 0.178 without processing, and higher than that of speech enhancement, speech separation and detector-gated muting at every speech level. Whisper recovers 5.2\% of the spoken words, compared with 89.0\% without processing and a chance floor of 2.2\%. On the real recordings, the speech proportion estimated by a voice activity detector falls from 8.5\% to 2.7\%. The method reduces, but does not eliminate, spoken content and speaker identity.
\end{abstract}


\end{frontmatter}

\section{Introduction}
\label{sec:introduction}

The number of people aged 65 and over is projected to reach 2.2 billion by the late 2070s, exceeding the number of children under 18 \cite{un2024wpp}. Population ageing is uneven across regions and most advanced in Europe and East Asia; where it is advanced, supporting older adults to live independently and safely in their own homes has become a pressing concern. Activity tracking technologies can help by monitoring daily routines unobtrusively, detecting health issues and enabling timely assistance \cite{kim2022home}. Among the available modalities, acoustic sensing is low in cost and captures rich contextual information without wearable devices or extensive infrastructure \cite{navarro2018real}: sounds associated with routines such as cooking, washing or moving around the home can reveal activity patterns and anomalies \cite{nicolaou2022tracking}.

Unlike cameras, which raise well-documented privacy concerns \cite{mujirishvili2023acceptance, demiris2009older}, or wearables, which require user compliance and regular charging \cite{hvalivc2022benefits}, microphones can be integrated into existing home infrastructure and operate without the occupants' involvement. However, microphones bring privacy risks of their own. An always-on microphone can capture private conversations \cite{iravantchi2021privacymic, boovaraghavan2024kirigami}, and speech can reveal the identity of the speaker \cite{qian2018hidebehind, nelus2021privacy}. In our deployment of an acoustic sensing system in care homes (Section~\ref{sec:motivation}), the most important issue raised by residents and their carers was that the system could record their conversations; similar concerns about always-listening microphones are well documented for smart speakers \cite{lau2018alexa, malkin2019privacy}. A second practical barrier is data: acoustic activity recognition relies on machine learning models trained on labelled sounds, and labelled recordings from real homes are costly to collect and can be gathered only over short periods.

In this paper we present a speech-removal front end for acoustic sensing systems: a processing stage that sits between the microphone and every downstream analysis and passes on audio from which speech has been suppressed. The aim of the system is to achieve two goals: (i) remove speech from the captured audio and (ii) keep the sounds needed to recognise the activities taking place. The privacy we target is the spoken content and the identity of the speaker; we measure both, and we state explicitly which other inferences from audio are outside the scope of this work (Section~\ref{sec:threatmodel}). Because in-home speech cannot be collected for training without the privacy risk we aim to remove, the model is trained on synthetic mixtures of environmental sounds and public speech. The speech-removal model is a convolutional U-Net that estimates a mask on the short-time Fourier transform (STFT) of the input and returns a waveform. Activity recognition is performed with VGGish \cite{hershey2017cnn}, a convolutional network pre-trained on a large corpus of YouTube videos, followed by a shallow classifier.

We evaluate the approach with group-disjoint cross-validation on mixtures of environmental sounds from ESC-50 \cite{piczak2015esc} and domestic recordings from SINS \cite{dekkers2017sins} with speech from LibriSpeech \cite{panayotov2015librispeech}, on speech from corpora not seen in training and on real home recordings from our AudioHive study, in which participants recorded their routines with a smartphone app we developed \cite{audiohive}. Privacy is measured as the detectability of speech by voice activity detectors (VADs) such as Silero VAD \cite{SileroVAD}, the recoverability of its content by Whisper speech recognition \cite{radford2023whisper} and the identifiability of the speaker. On both ESC-50 and SINS, the model preserves more activity information than VAD-gated muting and the speech enhancement and separation baselines at every signal-to-noise ratio (SNR), while Whisper recovers 4.7-7.6\% of the spoken words after removal and the equal error rate (EER) of a speaker-verification attack rises to 0.40-0.45; these results carry over to speech from corpora not seen in training, and on the real recordings speech removal lowers the VAD-estimated speech proportion from 8.5\% to 2.7\% without a significant change in activity recognition.

This paper makes three contributions:
\begin{enumerate}
    \item We formulate speech privacy in in-home acoustic sensing as a speech-removal front end, motivated by the constraints of a real assisted-living deployment, and define what it protects.
    \item We present a speech-removal model that masks the STFT of the input and returns a waveform, trained on synthetic speech-background mixtures, avoiding the need for privacy-sensitive in-home speech recordings.
    \item We evaluate the privacy-utility trade-off against VAD-gated muting, speech enhancement and separation models and a reconstruction-only control, on synthetic mixtures, unseen speech corpora and real home recordings.
\end{enumerate}

The rest of the paper is organised as follows: Section~\ref{sec:relatedwork} reviews related work, Section~\ref{sec:motivation} derives the requirements, Sections~\ref{sec:data} and \ref{sec:methods} describe the data and methods, Section~\ref{sec:results} reports the results, Section~\ref{sec:discussion} discusses them and Section~\ref{sec:conclusions} concludes the paper.

\section{Related Work}
\label{sec:relatedwork}

\subsection{Acoustic activity recognition in the home}

Acoustic sensing has been deployed in industrial monitoring, where microphones detect anomalous machine sounds \cite{huang2022sensing}; in healthcare, including respiratory monitoring, fall detection and sleep assessment \cite{kim2018detection, wang2020elderly, mallegni2022sensing}; and in smart cities for urban noise monitoring \cite{bello2019sonyc} and, with fibre-optic distributed acoustic sensing, traffic monitoring \cite{ye2023traffic}. In the home, environmental sounds can identify activities such as cooking, dishwashing and vacuum cleaning \cite{dekkers2018dcase}, and activity recognition in smart homes supports applications in assisted living and healthcare \cite{benmansour2015multioccupant}. Public datasets from real homes remain scarce, owing to privacy concerns, the cost of labelling and the variability of home environments. One of the few is SINS \cite{dekkers2017sins} (Section~\ref{sec:backgrounds}).

The Detection and Classification of Acoustic Scenes and Events (DCASE) challenges have advanced the field since 2013 through shared tasks, datasets and evaluation protocols \cite{stowell2015detection, mesaros2018dcase}, with Task~1 (acoustic scene classification) and Task~4 (sound event detection in domestic environments) of particular relevance \cite{schmid2024dcase, cornell2024dcase}. These challenges have also shown that models trained in one acoustic environment degrade when recording conditions change, for example when audio is captured with different devices \cite{heittola2020acoustic}, a problem that is acute in homes, where each environment is different. Transfer learning from large audio corpora addresses both this mismatch and the scarcity of labels: embeddings from models pre-trained on large audio corpora, such as VGGish \cite{hershey2017cnn}, trained on YouTube videos, and the pre-trained audio neural networks (PANNs) \cite{kong2020panns}, trained on AudioSet \cite{gemmeke2017audio}, allow classifiers to be trained from a small number of labelled examples from the target environment \cite{kong2020panns}.

\subsection{Speech separation and enhancement}

Speech separation isolates individual sources in a mixture, most prominently the speakers in the cocktail-party problem \cite{agrawal2023review}. The standard benchmark is WSJ0-2mix \cite{hershey2016deep}, two-speaker mixtures of Wall Street Journal utterances, on which models are compared by the improvement in scale-invariant signal-to-noise ratio (SI-SNRi) over the mixture. Transformer and state-space models dominate: SepFormer \cite{subakan2021attention} introduced self-attention for long-range dependencies and SepReformer \cite{shin2024separate} an asymmetric encoder-decoder; SPMamba \cite{li2024spmamba} uses state-space layers, and MossFormer2 \cite{zhao2024mossformer2} reaches 24.1~dB SI-SNRi on WSJ0-2mix. A recent survey reviews these advances \cite{araki2025sourcesep}. More realistic benchmarks have followed: LibriMix \cite{cosentino2020librimix}, built from LibriSpeech \cite{panayotov2015librispeech}; WHAM! \cite{wichern2019wham}, which adds real noise, and WHAMR! \cite{maciejewski2020whamr}, which adds reverberation as well; and the 20,000-hour LibriheavyMix \cite{jin2024libriheavymix}. These models are optimised to separate speakers from each other, not speech from the diverse environmental sounds of a home \cite{kavalerov2019universal}.

Speech enhancement removes background noise to improve speech quality. The Facebook Denoiser \cite{defossez2020real}, based on the DEMUCS architecture, maps noisy waveforms to clean speech with a convolutional encoder-decoder and skip connections. Later work followed three directions: attention-based time-frequency models such as CMGAN \cite{cao2022cmgan}, MP-SENet \cite{lu2023mpsenet} and TF-GridNet \cite{wang2023tfgridnet}; score-based diffusion models such as SGMSE+ \cite{richter2023speech}; and state-space models with a lower computational cost such as SEMamba \cite{chao2024semamba}. Across these directions the objective is to keep speech and discard the background \cite{braun2020data}. The underlying architectures, however, do not depend on which source is kept, which suggests that they can be used for the inverse task. We examine that inverse task empirically, and we include the Facebook Denoiser and SepFormer as baselines.

\subsection{Privacy-preserving acoustic sensing}
\label{sec:privacysensing}

A body of work limits what an audio sensor can reveal about speech. One line of research moves sensing outside the audible band: PrivacyMic \cite{iravantchi2021privacymic} recognises daily activities from ultrasonic and infrasonic frequencies only, so intelligible speech is never recorded. A second line degrades the signal until speech is unintelligible. SAMoSA \cite{mollyn2022samosa} subsamples smartwatch audio to 1~kHz or less and combines it with motion data. For long-term conversation analysis, Nespoli et al. \cite{nespoli2023longterm} combine spectral smoothing with speaker anonymisation by the McAdams coefficient, and Pohlhausen et al. \cite{pohlhausen2024longterm} add temporal smoothing and very low sampling rates; both measure privacy with speech and speaker recognition. Wyatt et al. \cite{wyatt2007conversation} detect conversations and speaker turns from features chosen to be insufficient for reconstructing speech, but assume that every conversant wears a recording device. A third line keeps raw audio on the device and releases only screened representations: Nelus and Martin \cite{nelus2021privacy} learn features that support sound classification while inhibiting speaker recognition, and Qian et al. \cite{qian2018hidebehind} conceal the speaker's voiceprint by voice conversion before speech is sent for recognition. Finally, a fourth line acts on speech segments. The TILES Audio Recorder \cite{feng2018tiles} uses a VAD to store features only while the wearer speaks; VAD-gated capture, the inverse, switches recording off whenever speech is detected. Kirigami \cite{boovaraghavan2024kirigami} showed that fine-tuned speech recognition models can recover speech from common audio featurisations (an average word error rate (WER) of 44.43\%) and proposed a lightweight filter that removes speech segments before activity recognition.

Most of these approaches discard part of the signal (band restriction, subsampling, gating or removal of whole segments) or restrict analysis to purpose-built features, while voice conversion and anonymisation alter the voice but keep its content. In contrast, the approach presented in this paper suppresses speech within each time-frequency region and returns a waveform, so background sounds that overlap with speech are attenuated rather than discarded and standard audio analyses remain applicable. Following Kirigami, we measure privacy with speech recognition, and following Nespoli et al. and Pohlhausen et al., also with speaker verification.

\section{Motivation and requirements}
\label{sec:motivation}

This work is motivated by our experience in the ADAPTIVE project (AI-based Assistive and Passive Technology for non-Invasive Elderly care) \cite{ukri}. The project, a collaboration between the University of Kent, the smart-home technology provider MiiCare \cite{miicare}, East Kent Hospitals University NHS Foundation Trust and Bristol City Council, deployed acoustic sensing in three care homes over 12 months to assess the gait, and hence the fall risk, of residents living with dementia. The deployment exposed three constraints. First, industry partners required systems that could be installed and made operational within weeks, so the data available in a new home were limited to a short collection window. Second, these data were unlabelled; labels were produced by human annotators listening to the recordings, which is slow and costly. Third, and most important, residents and carers were reluctant to consent because conversations could be captured at any stage. The project's solution was to switch off audio capture whenever speech was detected. However, this approach can lead to sub-optimal performance: when residents have a radio playing in the background, for example, the speech detector may trigger frequently, causing the system to miss major parts of their daily activities.

From this experience we derive the requirements for the system:
\begin{enumerate}
    \item The system should remove speech from the audio before it is stored or analysed, for regular occupants and for visitors.
    \item The system should keep the sounds needed for acoustic activity recognition, including those that overlap with speech.
    \item The system should operate with a limited number of labelled examples from the target home.
    \item The system should support rapid deployment: it should not require speech recordings from the home for training, and it should need only a short data collection period.
\end{enumerate}

\subsection{Threat model}
\label{sec:threatmodel}

Figure~\ref{fig:propose} places the speech-removal front end between the microphone and every downstream analysis. We assume an adversary who obtains the audio after this stage, for example from storage or from a remote service, and who applies off-the-shelf models to it. We consider three questions: whether speech can still be detected, whether its spoken content can be recovered and whether the speaker can be identified. We measure the first with three VADs, the second with automatic speech recognition (ASR) and the third with a speaker-verification attack in which the adversary holds enrolment recordings of the occupants (Section~\ref{sec:privacy-eval}). Other inferences are outside the scope of this work: the speaker's gender, age, language or health from residual voice characteristics, and information carried by non-speech sounds, such as the occupants' routines, which activity recognition deliberately keeps. We also do not consider an adversary who fine-tunes recognition models on processed audio, which Kirigami \cite{boovaraghavan2024kirigami} showed can recover more speech than off-the-shelf models.

\begin{figure}[t]
\centering
\begin{tikzpicture}[
  node distance=5mm and 7mm,
  box/.style={draw, rounded corners=1pt, align=center, font=\small, minimum height=10mm, inner sep=4pt},
  fw/.style={box, fill=black!8, very thick},
  arr/.style={-{Latex[length=2mm]}, thick},
  lbl/.style={font=\footnotesize, align=left}]
  \node[box] (mic) {Microphone\\[-1pt]{\footnotesize speech + background}};
  \node[fw, right=of mic] (fw) {Speech-removal\\front end};
  \node[box, right=of fw] (down) {Stored audio and\\downstream analyses\\[-1pt]{\footnotesize activity recognition, \dots}};
  \draw[arr] (mic) -- (fw);
  \draw[arr] (fw) -- (down);
  \node[lbl, anchor=north west] (in) at ([yshift=-9mm]mic.west |- down.south) {\textbf{Measured in this paper}\\
    speech detectability: three VADs\\
    spoken content: ASR (Whisper)\\
    speaker identity: speaker verification};
  \node[lbl, anchor=north east] (out) at ([yshift=-9mm]down.south east) {\textbf{Outside the scope}\\
    gender, age, language, health\\
    information in non-speech sounds\\
    adversaries who fine-tune on output};
  \draw[dashed, thin] (down.south) -- ++(0,-4mm) coordinate (a) -| (in.north);
  \draw[dashed, thin] (a) -| (out.north);
\end{tikzpicture}
\caption{The speech-removal front end sits between the microphone and all storage and analysis. An adversary with access to its output is assumed to apply off-the-shelf models; the threats measured in this paper and those outside its scope are listed below the pipeline.}
\label{fig:propose}
\end{figure}

\section{Data}
\label{sec:data}

We use two kinds of audio: background sounds that carry activity labels, from ESC-50, SINS and our AudioHive study, and speech, from LibriSpeech and, for testing only, from VCTK and Common Voice. Speech is placed in rooms with measured or simulated room impulse responses (RIRs). We resample all audio to 16~kHz.

\subsection{Background sounds}
\label{sec:backgrounds}

ESC-50 \cite{piczak2015esc} contains 2,000 five-second clips of environmental sounds in 50 classes of 40 clips each, distributed at 44.1~kHz. Its five official folds keep clips cut from the same source recording together, with the exception of four recordings (nine clips) that span two folds; we use the official folds for cross-validation.

SINS \cite{dekkers2017sins} was recorded while one person lived alone in a vacation home for one week, with 13 arrays of four microphones distributed over the rooms. We use its derivative from DCASE 2018 Task 5 \cite{dekkers2018dcase}: four-channel ten-second segments from the four arrays in the combined living room and kitchen, labelled with nine activities (absence, cooking, dishwashing, eating, other, social activity, vacuum cleaning, watching TV and working). We average the four channels of each segment. A session is one continuous activity, for example a full cooking session, recorded by all four arrays at the same time; the four official folds keep each session within one fold, so the same event never appears in both training and test data. For evaluation we draw at most 75 segments per class and fold, 2,643 in total, each from one randomly chosen array, so that the same moment is not evaluated four times. Two activities, social activity and watching TV, contain speech in the background itself, so a speech-removal front end is expected to affect them. The room acoustics of the SINS home (reverberation time, direct-to-reverberant ratio) are not documented.

\subsection{Speech}
\label{sec:speech}

LibriSpeech \cite{panayotov2015librispeech} provides read English speech in three speaker-disjoint subsets: train-clean-100 (251 speakers, 100.6~h) trains the speech-removal models, dev-clean (40 speakers) selects their checkpoints and test-clean (40 speakers, 20 female and 20 male) is used only for testing. No test speaker is therefore heard during training.

To test speech from other corpora, which the models never see during training, we use two further sources. VCTK version 0.92 \cite{yamagishi2019vctk} contains read speech from 110 English speakers with a range of accents, recorded in a studio; we use the 109 speakers with transcripts (63 female, 46 male; 41.5~h). Common Voice \cite{ardila2020commonvoice} is crowd-sourced on participants' own devices; we use the validated English test set of the 2017 release (3,995 clips, 5.0~h), which covers many microphones and accents but has no speaker identities. Finally, two-talker mixtures place one female and one male test-clean speaker in the same mixture, to test conversations rather than a single speaker.

\subsection{Room impulse responses}
\label{sec:rirs}

We convolve speech with RIRs from the RIRS\_NOISES collection \cite{ko2017rir}. We use its 60,000 simulated RIRs from 600 rooms for training (90\% of the rooms) and validation (the remaining 10\%), and its 325 measured RIRs, from the AIR, RWCP and REVERB 2014 sets, only for testing; the collection's RIR list groups the measured RIRs into 93 rooms.

\subsection{AudioHive}
\label{sec:audiohive}

AudioHive is a data-collection study we conducted with a purpose-built smartphone app for Android (Java) and iOS (Swift) \cite{audiohive}. Participants recorded their daily routines with the app, with the phone placed on the table nearest to the activity, and marked the start of each activity by tapping its label. In the first phase participants were asked not to speak; in the second they were asked to speak whenever their activities allowed, for example to discuss non-personal topics, describe what they were doing or read from a provided script.

The Android app stored audio in Advanced Audio Coding (AAC) format at 8~kHz and the iOS app at 12~kHz. Each tap marks the start of an activity, which lasts until the next tap or the end of the recording. We harmonised spelling and tense variants of the activity names and kept the activities performed by at least three participants in both phases: brushing teeth, preparing breakfast, preparing coffee or tea and washing dishes. We cut each activity into ten-second windows; a final window shorter than ten seconds is kept if it lasts at least five seconds. After removing labelled recordings without an audio file (nine in phase 1 and 67 in phase 2), the analysis covers 1,295 windows from nine participants (one female, eight male) in phase 1 and 1,475 windows from seven participants (three female, four male) in phase 2. No participant identifier of the app appears in both phases. Because most phase-1 audio comes from the 8~kHz Android app and most phase-2 audio from the 12~kHz iOS app, we band-limit all AudioHive audio to 4~kHz, so that both phases share the same bandwidth.

\section{Methods}
\label{sec:methods}

\subsection{Overview}

Figure~\ref{fig:overview} summarises the evaluation pipeline. Test mixtures are built from a background clip and reverberant speech at a calibrated SNR. Every speech-removal method, the proposed model as well as each baseline, takes the mixture waveform and returns a waveform of the same length. Every output then passes through the same activity classifier, the same three VADs, the same speech recogniser and the same speaker-verification model, so that all methods are compared under one pipeline. All models are trained and evaluated under a single protocol with group-disjoint cross-validation: one speech-removal model and one activity classifier are trained per test fold, and no background recording, SINS activity session or speaker is shared between training and test data, apart from the four ESC-50 recordings noted in Section~\ref{sec:backgrounds}.

\begin{figure}[t]
\centering
\begin{tikzpicture}[
  node distance=4mm and 6mm,
  box/.style={draw, rounded corners=1pt, align=center, font=\footnotesize, minimum height=9mm, inner sep=3pt},
  main/.style={box, fill=black!8, very thick},
  arr/.style={-{Latex[length=1.8mm]}, thick}]
  \node[box] (bg) {Background\\ESC-50 / SINS};
  \node[box, right=6mm of bg] (mix) {Mixing at\\SNR $-5$ to 15 dB\\+ speech mask};
  \node[box, below=of mix] (rir) {Measured\\RIR};
  \node[box, left=6mm of rir] (sp) {Speech\\LibriSpeech, VCTK,\\Common Voice};
  \node[main, right=of mix] (rm) {Speech removal\\proposed U-Net\\or baseline};
  \node[box, right=of rm, yshift=11mm] (aar) {VGGish + SVM\\per-class F1};
  \node[box, right=of rm] (vad) {3 VADs vs\\speech mask};
  \node[box, right=of rm, yshift=-11mm] (asr) {Whisper hit rate,\\speaker EER};
  \draw[arr] (bg) -- (mix);
  \draw[arr] (sp) -- (rir);
  \draw[arr] (rir) -- (mix);
  \draw[arr] (mix) -- (rm);
  \draw[arr] (rm.east) -- ++(2mm,0) |- (aar.west);
  \draw[arr] (rm.east) -- (vad.west);
  \draw[arr] (rm.east) -- ++(2mm,0) |- (asr.west);
\end{tikzpicture}
\caption{Evaluation pipeline. Each speech-removal method returns a waveform, which is evaluated for utility (activity recognition) and privacy (speech detectability, spoken content and speaker identity) by the same models.}
\label{fig:overview}
\end{figure}

\subsection{Test mixtures}
\label{sec:mixing}

Each test mixture combines one background clip (five seconds for ESC-50, a ten-second segment for SINS) with speech from one test speaker. We assign speakers to clips in turn, from an order that alternates female and male speakers. The speech track is built from whole utterances, which are never cropped: after a random initial offset of up to 1~s, utterances drawn at random are placed one after another, separated by pauses of 0.5-2~s, until no remaining utterance fits. Because most LibriSpeech utterances are long, 98.9\% of the five-second ESC-50 mixtures contain a single utterance, whereas 63.8\% of the ten-second SINS mixtures contain two or three. The track is convolved with a measured RIR, aligned on its direct-path peak.

A ground-truth speech mask marks every 10~ms frame that lies inside an utterance and whose energy in the dry speech is within 30~dB of that utterance's loudest frame. The speech level is the mean power over these frames, and the background level the mean power over frames within 50~dB of the clip's loudest frame, which excludes the digital silence that pads many ESC-50 clips. Because the background level is averaged over its active frames, the local SNR of short or impulsive events varies within a clip; Section~\ref{sec:res-utility} shows the effect on such classes. The speech is scaled to SNRs of $-5$, 0, 5, 10 and 15~dB, with the same speech track used at every SNR so that results are paired across SNRs. Measurements of everyday listening situations report mostly positive SNRs, with speech typically 5 to 15~dB above the background \cite{pearsons1977speech} and few situations below 0~dB \cite{smeds2015snr}; our range covers these conditions and two harder ones. If the mixture would clip, mixture, background and speech are scaled down together, which preserves the SNR. For every mixture we record the ground-truth speech proportion, the mean of the mask: it is 46.1\% on average for ESC-50 and 52.1\% for SINS. For comparison, Pohlhausen et al. \cite{pohlhausen2022nearear} labelled speech in 45.2\% of 1.98~h of everyday near-ear recordings, and estimated the wearer's own voice at 10.3\% of a day. The out-of-distribution sets (Section~\ref{sec:speech}) follow the same protocol; in two-talker mixtures, the two speakers are brought to equal speech power and convolved with two different RIRs measured in the same room where the room has more than one, and otherwise with the same RIR (17.8\% of the ESC-50 and 16.6\% of the SINS two-talker mixtures).

\subsection{Proposed speech-removal model}
\label{sec:model}

Let $x(t) = s(t) + b(t)$ be a mixture of reverberant speech $s$ and background $b$, and $X(f,\tau)$ its STFT (frame length of 512 samples, hop size of 128 samples, Hann window). The model estimates a mask $M(f,\tau) \in [0,1]$ from the log-magnitude of $X$ and returns the background estimate
\begin{equation}
\hat{b} = \mathrm{iSTFT}\left(M \odot X\right),
\end{equation}
where iSTFT denotes the inverse STFT. The output reuses the phase of the mixture, so no spectrogram inversion or phase reconstruction is needed. The input is divided by its standard deviation and the output is multiplied by it, which makes the model independent of the recording level. The mask is predicted, from the batch-normalised log-magnitude, by a 2-D U-Net \cite{ronneberger2015u} with four levels of 32, 64, 128 and 256 channels and a bottleneck of 512 channels. Each level applies two $3\times3$ convolutions with batch normalisation and LeakyReLU activations; the encoder downsamples with $2\times2$ max-pooling, the decoder upsamples with transposed convolutions and receives skip connections from the encoder, and a $1\times1$ convolution with a sigmoid produces the mask.

The model is trained to reconstruct the background waveform with
\begin{equation}
\mathcal{L} = \frac{1}{T}\lVert b - \hat{b} \rVert_1 + \lambda \, \frac{1}{K} \sum_{k=1}^{K} \left( \mathcal{L}_{\mathrm{sc}}^{(k)} + \mathcal{L}_{\mathrm{mag}}^{(k)} \right), \qquad \lambda = 0.5,
\end{equation}
where $T$ is the number of samples and the second term is the multi-resolution STFT (MR-STFT) loss \cite{yamamoto2020parallel}: $\mathcal{L}_{\mathrm{sc}}^{(k)}$ is the spectral convergence and $\mathcal{L}_{\mathrm{mag}}^{(k)}$ the L1 distance between log-magnitudes at resolution $k$, with frame lengths of 512, 1024 and 2048 samples and 75\% overlap ($K=3$). Both terms are computed on signals divided by the standard deviation of the mixture. To examine the contribution of each term, we also train the ESC-50 models with the waveform L1 term alone and with the multi-resolution STFT term alone.

We generate training mixtures on the fly. Each item is a four-second crop of a background from the training folds. With probability 0.8, speech is added: whole utterances of a random train-clean-100 speaker are placed with pauses in a 20-second track, convolved with a simulated RIR and cropped to four seconds, since most training utterances are longer than four seconds. The SNR is drawn uniformly between $-10$ and 20~dB. We train each model for 25,000 steps with a batch size of 16, AdamW with a learning rate of $3 \times 10^{-4}$ and cosine decay, mixed precision and gradient clipping, which takes 1.3~h (median) on one NVIDIA RTX 2080 Ti. Every 1,000 steps the model is evaluated on a fixed validation set of 512 mixtures built from a held-out group of the training folds (10\% of the ESC-50 source recordings or of the SINS sessions), dev-clean speech and held-out simulated rooms. We keep the checkpoint with the highest scale-invariant signal-to-distortion ratio (SI-SDR) \cite{leroux2019sdr} between the estimated and the true background. The same criterion is used for every loss variant, so the comparison of losses is not affected by their different scales.

For AudioHive, one model is trained in the same way on the band-limited phase-1 windows, recorded with the instruction not to speak, with speech band-limited to 4~kHz; one phase-1 participant is held out for checkpoint selection. The model therefore needs no speech recorded in the homes, only speech-free recordings, here from other homes recorded with the same app.

\subsection{Baselines}
\label{sec:baselines}

We compare the proposed model with six methods, all of which return a waveform. \textbf{No processing} passes the mixture on unchanged and gives the lower bound for privacy; the \textbf{clean background} gives the upper bound for utility. \textbf{VAD-gated muting}, the solution used in our deployment, mutes every window in which Silero VAD \cite{SileroVAD} reports a speech probability above 0.5, with a 100~ms hangover and 10~ms fades. The \textbf{mel + Griffin-Lim control} converts the mixture to a 64-band mel spectrogram (25~ms window, hop size of 10~ms) and back, by a pseudo-inverse of the mel filterbank and 32 iterations of Griffin-Lim \cite{griffin1984signal}, without any speech removal; it shows whether reconstruction artefacts alone change what the detectors and the classifier see. Finally, the \textbf{Facebook Denoiser} \cite{defossez2020real} (pre-trained \texttt{dns64}) and \textbf{SepFormer} \cite{subakan2021attention} (\texttt{sepformer-wham16k-enhancement} \cite{speechbrain}) estimate the speech, which we subtract from the mixture after a least-squares projection onto it; the projection is needed because SepFormer is trained with a scale-invariant objective and its output has arbitrary level and sign.

\subsection{Utility evaluation}
\label{sec:utility-eval}

We intentionally use a lightweight VGGish/SVM classifier as a downstream utility probe; the focus of this paper is the speech-removal front end rather than advancing acoustic activity recognition. We first scale each clip to a fixed level (an active power of $-30$~dB relative to full scale), because the features depend on the input gain and the test mixtures carry an SNR-dependent scaling to avoid clipping. VGGish \cite{hershey2017cnn}, in its torchaudio implementation \cite{yang2022torchaudio}, computes a 64-band log-mel spectrogram (25~ms window, hop size of 10~ms) and returns one 128-dimensional embedding for every non-overlapping 0.96-second example. A clip is represented by the mean and standard deviation of its embeddings. After standardisation, a support vector machine (SVM) with a radial basis function kernel is trained \cite{scikit-learn}; $C \in \{0.1, 1, 10, 100\}$ and $\gamma$ are selected by leave-one-fold-out cross-validation within the training folds, maximising macro-F1. The classifier is trained on the clean backgrounds of the training folds and applied to the output of every method on the test fold. We report accuracy and per-class and macro precision, recall and F1, all computed from saved per-clip predictions, together with confusion matrices, which are released with the results.

To check whether the classifier assigns the same label to successive windows of one activity, we classify up to 30 consecutive ten-second segments of each SINS test session, from one array, and report how often successive windows receive the same prediction.

For AudioHive, the classifier is trained on the phase-1 windows, with its parameters selected by cross-validation grouped by participant. The effect of speech removal is measured within phase 2: every phase-2 window is classified unprocessed and after each method, which gives paired comparisons on the same windows. The accuracy on phase 1 under leave-one-participant-out cross-validation serves as the speech-free reference. Because no participant took part in both phases and the phases were mostly recorded on different devices, the difference between phase-1 and phase-2 accuracy is not attributed to speech.

\subsection{Privacy evaluation}
\label{sec:privacy-eval}

\textbf{Speech detectability.} We use three VADs of different design: Silero VAD \cite{SileroVAD}, a neural model that processes 512-sample windows (32~ms at 16~kHz), with a threshold of 0.5; the WebRTC VAD \cite{webrtcvad}, a statistical model applied to 30~ms frames in its most aggressive mode; and the SpeechBrain VAD \cite{speechbrain}, a convolutional recurrent deep neural network (CRDNN) trained on LibriParty (\texttt{vad-crdnn-libriparty}), which outputs a speech probability every 10~ms, with a threshold of 0.5. No minimum-duration rule is applied. Every decision is mapped onto the 10~ms frames of the ground-truth mask, and we report three quantities per VAD: the detected speech proportion, which is an estimate of how much of the audio contains speech and not a classification score; the detection rate, the share of ground-truth speech frames flagged as speech; and the false-alarm rate on non-speech frames. As ceilings, each VAD is also applied to the reverberant speech track alone and to the clean background alone, where any detection is a false alarm.

\textbf{Spoken content.} Whisper large-v3-turbo \cite{whisperturbo}, a model of the Whisper family \cite{radford2023whisper}, transcribes a fixed random subset of 400 test clips per dataset, the same clips for every method and SNR, in English with greedy decoding. Transcripts and references are normalised with Whisper's English text normaliser. We report the word hit rate, the share of reference words that Whisper recognises correctly, rather than the WER: Whisper produces words on speech-free audio, and such hallucinations inflate the WER without revealing any speech, whereas the hit rate counts only recovered words. The pooled WER of every condition is released with the results. As ceilings, Whisper is also run on the speech track alone and on the clean background, which also gives the floor of the hit rate: the share of reference words matched by chance in transcripts of speech-free audio.

\textbf{Speaker identity.} A speaker-verification attack uses ECAPA-TDNN embeddings \cite{desplanques2020ecapa} (SpeechBrain \texttt{spkrec-ecapa-voxceleb} \cite{speechbrain}). Each test speaker is enrolled with the mean embedding of five of their utterances that appear in no test mixture. Every output in the Whisper subset is scored, by cosine similarity, against all enrolled speakers; the trial against the mixture's own speaker is a target trial and the others are non-target trials. We report the EER, where 0.5 means that the speaker cannot be identified better than by chance. The attack is run for LibriSpeech, VCTK and the two-talker mixtures, in which both speakers give a target trial; Common Voice has no speaker labels.

\textbf{AudioHive.} No ground-truth speech mask or transcript exists for the real recordings. We therefore report each VAD's detected speech proportion before and after processing, and use Whisper's transcript of each unprocessed phase-2 window as the reference: we report how many words remain after each method and the share of reference words still recognised.

\subsection{Statistical analysis}
\label{sec:stats}

We report results pooled over folds; the standard deviation across folds and 95\% bootstrap confidence intervals from 1,000 clip-level resamples are given in the text for the main comparisons and for every condition in the released results. The proposed model is compared with each other method with the exact McNemar test \cite{mcnemar1947note} on per-clip classification correctness and with the Wilcoxon signed-rank test \cite{wilcoxon1945individual} on per-clip privacy scores, with Holm correction \cite{holm1979simple} within each family of tests. We break results down by SNR, activity class and speaker gender. Every number in this paper is produced by a script that saves per-sample predictions, the configuration, the random seed where one is used and the code version (see Data availability).

\section{Results}
\label{sec:results}

Throughout this section, the SNR is the level of the speech relative to the background: higher values mean louder speech. Audio examples of every method, drawn at random from the ESC-50 test mixtures together with their transcripts and predictions, will be released with the code (see Data availability).

\subsection{Activity recognition}
\label{sec:res-utility}

Table~\ref{tab:utility} reports macro-F1 for every method. On ESC-50, the proposed model preserves the most activity information at every SNR: macro-F1 falls from 0.636 at $-5$~dB to 0.479 at 15~dB, compared with 0.401 to 0.099 without processing and a clean-background ceiling of 0.685. The Facebook Denoiser is the best baseline from $-5$ to 10~dB (0.542 to 0.364); at 15~dB, VAD-gated muting is slightly higher (0.306 compared with 0.288). VAD-gated muting changes least with the speech level (0.425 to 0.306), but stays below the proposed model at every SNR, because muting also removes the background whenever speech is detected. The Griffin-Lim control stays close to the unprocessed mixture (0.409 compared with 0.401 at $-5$~dB, and 0.092 compared with 0.099 at 15~dB), so mel reconstruction alone does not reproduce the effect of speech removal on the classifier. SepFormer is below the unprocessed mixture at $-5$~dB (0.380 compared with 0.401) and above it at higher SNRs. The proposed model is above every baseline at every SNR (McNemar, Holm-corrected $p < 10^{-17}$); at 5~dB its macro-F1 of 0.574 has a 95\% bootstrap confidence interval of 0.550-0.591. The standard deviation of macro-F1 across the five ESC-50 folds is between 0.002 and 0.041.

On SINS, the clean background reaches a macro-F1 of 0.798, and speech affects activity recognition more than on ESC-50: without processing, macro-F1 falls from 0.451 at $-5$~dB to 0.057 at 15~dB. The proposed model preserves 0.735 to 0.520 (0.661 at 5~dB, with a 95\% confidence interval of 0.644-0.677) and is again the best method at every SNR (McNemar, Holm-corrected $p < 10^{-48}$ against every baseline). The Facebook Denoiser is again the best baseline (0.601 to 0.170), and the Griffin-Lim control stays close to the unprocessed mixture (0.443 compared with 0.451 at $-5$~dB). The macro-F1 of VAD-gated muting is below that of the unprocessed mixture at every SNR and falls to 0.040 at 10 and 15~dB, where almost all of its outputs are classified as social activity or watching TV, the two activities with speech in the background. SepFormer is below the unprocessed mixture at $-5$ and 0~dB. The standard deviation of macro-F1 across the four SINS folds is between 0.003 and 0.047.

Per class, speech affects almost every ESC-50 activity: with speech 5~dB above the background, mean per-class F1 falls from 0.685 to 0.233. After speech removal it recovers to 0.574, and two of the 50 classes reach their clean-background F1. The largest remaining losses are for impulsive or broadband sounds and for non-speech vocal sounds: clock tick (0.37 compared with 0.68 clean), wood door knock (0.50 compared with 0.80), wind (0.23 compared with 0.49), thunderstorm (0.55 compared with 0.81), cat (0.25 compared with 0.48) and snoring (0.46 compared with 0.68). On SINS at 5~dB, speech removal raises macro-F1 from 0.178 to 0.661, but no class reaches its clean-background F1. Vacuum cleaning (0.98 compared with 1.00 clean) and watching TV (0.91 compared with 0.93) are almost unaffected, whereas dishwashing (0.48 compared with 0.73), eating (0.59 compared with 0.81) and cooking (0.69 compared with 0.90) lose most. Social activity, whose background itself contains speech, falls from 0.88 to 0.68, while watching TV, which also contains speech, is preserved. \ref{app:perclass} gives the per-class precision and recall for both datasets.

\textbf{Label consistency in SINS.} On the clean audio of 266 test sessions (3,939 windows, from one array per session and therefore a different sample from the evaluation segments of Table~\ref{tab:utility}), successive ten-second windows of the same session receive the same prediction in 67.1\% of cases, and on average 79.9\% of a session's windows agree with its majority prediction. A majority vote over each session is correct for 81.2\% of the sessions, compared with an accuracy of 69.9\% for single windows. Consistency is highest for vacuum cleaning (all successive windows agree), watching TV (88.9\%) and social activity (86.7\%), and lowest for working (53.7\%) and the heterogeneous class other (59.2\%). Window-level predictions are therefore stable for most activities, and a deployed system could aggregate them over an activity session; we report window-level results throughout.

\subsection{Privacy}
\label{sec:res-privacy}

Table~\ref{tab:privacy} reports the privacy results on ESC-50 and SINS. On ESC-50, the ceilings show what each measure finds when speech is certainly present: on the reverberant speech track alone, Silero VAD detects 96.0\% of the speech frames, Whisper recognises 96.3\% of the reference words and the speaker-verification EER is 0.040. On the clean background, Silero flags 0.5\% of the frames as speech and Whisper matches 2.2-2.4\% of the reference words by chance, which is the floor of the hit rate. Silero flags at least one frame in 294 of the 2,000 clean backgrounds, most often in classes of non-speech vocal and animal sounds (coughing 29, sneezing 27, insects 26, pig 20), which suggests false alarms of the detector rather than speech in the source recordings. We did not verify these detections by listening. The speaker EER of the clean background is 0.495, close to the 0.5 of chance.

The proposed model reduces all three measures towards their speech-free floors. Silero detects 1.4-2.5\% of the speech frames, compared with 58.3-94.4\% without processing; Whisper recovers 4.7-6.6\% of the reference words, compared with 69.3-95.1\%; and the speaker EER rises to 0.410-0.454, compared with 0.042-0.117. The reductions in Silero and CRDNN detection and in the word hit rate relative to the unprocessed mixture, the Griffin-Lim control, SepFormer and the Facebook Denoiser are significant at every SNR (Wilcoxon, Holm-corrected $p < 10^{-13}$); we did not test the EER for significance. Speaker identity is not removed completely: an EER of 0.41-0.45 is above the 0.04 of the speech track but below the 0.50 of chance. The CRDNN VAD agrees in ordering but remains more sensitive, flagging 11.1-13.0\% of the speech frames after removal, compared with 3.4\% of the same frames on the clean background. The WebRTC VAD flags 26.2\% of the frames of the clean backgrounds as speech, so it does not separate speech from these backgrounds, and on the synthetic mixtures we report only Silero and the CRDNN VAD.

The Griffin-Lim control answers whether reconstruction alone hides speech. It does not: its Silero detection rate (54.3-93.7\%), word hit rate (69.3-95.0\%) and speaker EER (0.041-0.120) are close to those of the unprocessed mixture at every SNR. Mel reconstruction therefore neither fools the detectors nor removes the spoken content. VAD-gated muting is scored by Silero, the detector that drives it, which explains its near-zero Silero rate; the CRDNN VAD, which does not drive the gate, detects 17.6\% of the speech frames in its output at $-5$~dB and 1.7-7.0\% at 0-15~dB, compared with 11.1-13.0\% for the proposed model. Judged by Whisper, its word hit rate (3.3-13.9\%) does not differ significantly from that of the proposed model at 0-15~dB, but is higher at $-5$~dB (13.9\% compared with 6.6\%, $p = 1.3 \times 10^{-7}$), when quieter speech escapes the gate; its speaker EER is 0.425-0.496. The enhancement and separation baselines leave far more speech: SepFormer 42.2-70.5\% and the Facebook Denoiser 19.9-28.7\% of the reference words.

On SINS, the speech-free floor of the detectors is higher, because two activities contain speech in the background itself: on the clean background, Silero flags 8.0-8.1\% of the ground-truth speech frames and the CRDNN VAD 14.8\%, and 464 of the 710 clean backgrounds in which Silero flags any frame are social activity or watching TV. On the speech track alone, Silero detects 96.4\% of the speech frames, Whisper recognises 97.3\% of the reference words and the speaker EER is 0.009. After removal, Silero detects 7.7-8.6\% of the speech frames, compared with 43.1-93.9\% without processing; this does not differ significantly from the clean background at $-5$ to 5~dB (Wilcoxon, Holm-corrected $p \geq 0.18$) and is slightly above it at 10 and 15~dB (8.6\% and 8.3\% compared with 8.1\% and 8.0\%). The CRDNN VAD flags 19.2-20.8\% of the speech frames, compared with 14.8\% on the clean background. Whisper recovers 5.0-7.6\% of the reference words, compared with 55.9-95.8\% without processing and a floor of 2.9-3.0\%, and the speaker EER rises to 0.404-0.422, compared with 0.012-0.152. VAD-gated muting leaks more words than the proposed model at $-5$ to 5~dB (22.3\%, 15.0\% and 9.1\% compared with 5.0\%, 7.0\% and 6.8\%; $p \leq 0.04$) and does not differ significantly at 10 and 15~dB. As on ESC-50, the Griffin-Lim control is close to the unprocessed mixture (word hit rate 56.1-95.6\%), and SepFormer (26.0-70.7\%) and the Facebook Denoiser (9.1-33.6\%) leave more words than the proposed model.

\subsection{Loss ablation}
\label{sec:res-ablation}

The loss terms trade utility against privacy. For activity recognition, the combined loss and the multi-resolution STFT loss alone do not differ significantly except at 5~dB (macro-F1 0.574 compared with 0.553; McNemar, Holm-corrected $p = 0.012$), and the combined loss exceeds the L1 loss alone at 5-15~dB (e.g. 0.479 compared with 0.385 at 15~dB, $p < 10^{-6}$; Table~\ref{tab:utility}). For privacy the order reverses: the L1 loss alone leaves the least speech, with a word hit rate of 3.1-3.8\% compared with 4.7-6.6\% for the combined loss and 6.6-7.8\% for the multi-resolution STFT loss alone, and a speaker EER of 0.450-0.478 (Table~\ref{tab:privacy}). The L1-only variant also reached the highest validation SI-SDR on all five folds. A waveform loss alone thus suppresses more of the speech but also more of the background detail that distinguishes activities, whereas the spectral term preserves that detail. We keep the combined loss as the proposed model because it preserves the most activity information while recovering only 1.4-2.7 percentage points more reference words than the L1-only variant.

\subsection{Speech from unseen corpora}
\label{sec:res-ood}

Table~\ref{tab:ood} compares the four speech sources on the ESC-50 and SINS backgrounds at 5~dB, with the models trained on LibriSpeech train-clean-100. The sets contain different amounts of speech: the ground-truth speech proportion is 46.1\% with LibriSpeech, 39.0\% with VCTK, 37.8\% with Common Voice and 62.1\% with two talkers on ESC-50, and 52.1\%, 42.2\%, 25.1\% and 72.0\% on SINS. The unprocessed mixtures therefore affect activity recognition to different degrees, and we compare the methods within each set rather than across sets.

On ESC-50, the gains in activity recognition carry over to the unseen corpora. The proposed model preserves the most activity information on every set and at every SNR (McNemar, Holm-corrected $p < 10^{-8}$ against every baseline): from $-5$ to 15~dB, macro-F1 falls from 0.654 to 0.553 on VCTK and from 0.629 to 0.499 on Common Voice, compared with 0.453 to 0.167 and 0.457 to 0.168 without processing. Two talkers are the hardest case: macro-F1 falls from 0.583 to 0.362, compared with 0.352 to 0.049 without processing and 0.347 to 0.181 for VAD-gated muting.

The privacy results also carry over. After removal, Silero detects 0.9-2.0\% of the VCTK and 1.8-4.8\% of the Common Voice speech frames, compared with 55.5-81.2\% and 55.1-92.9\% without processing; on the VCTK speech track alone it detects 82.0\%, compared with 96.0\% for LibriSpeech. Whisper recognises 4.7-6.5\% and 6.4-7.4\% of the reference words, compared with 69.5-95.7\% and 60.3-89.8\% without processing and hit-rate floors of 2.0-2.4\% and 3.0-3.2\% on the clean background. On VCTK, the speaker EER rises to 0.473-0.485, compared with 0.067-0.191 without processing, which is closer to chance than on LibriSpeech (0.410-0.454). As on LibriSpeech, VAD-gated muting leaves more words than the proposed model at $-5$~dB (10.2\% compared with 5.1\% on VCTK and 11.8\% compared with 6.4\% on Common Voice; Wilcoxon, Holm-corrected $p < 10^{-3}$) and does not differ significantly at 0-15~dB.

With two talkers, the reference contains the words of both, and Whisper recognises only 52.7\% of them even on the speech track alone; the speaker EER of the speech track alone is 0.275, as each embedding is computed from two overlapping voices. After removal, more speech remains detectable than with one talker: Silero detects 2.0-9.1\% and the CRDNN VAD 18.0-25.8\% of the speech frames, both increasing with the SNR, compared with 1.4-2.5\% and 11.1-13.0\% with one talker. Whisper recognises 4.3-4.8\% of the reference words, compared with 30.1-48.8\% without processing, and the speaker EER is 0.435-0.444, compared with 0.288-0.315. VAD-gated muting leaves fewer recognised words than the proposed model at 10 and 15~dB (2.7\% and 2.6\% compared with 4.5\% and 4.6\%; $p = 0.011$ and $p = 0.008$), more at $-5$~dB (6.7\% compared with 4.3\%, $p < 10^{-3}$), and preserves much less activity information.

On SINS, the same pattern holds. The proposed model again preserves the most activity information on every set and at every SNR (McNemar, Holm-corrected $p < 10^{-21}$ against every baseline), with macro-F1 of 0.759 to 0.611 on VCTK, 0.768 to 0.629 on Common Voice and 0.698 to 0.275 with two talkers from $-5$ to 15~dB, compared with 0.512 to 0.089, 0.633 to 0.246 and 0.344 to 0.044 without processing. Whisper recovers 6.6-8.7\% of the VCTK and 7.8-12.0\% of the Common Voice reference words after removal, compared with 56.6-96.6\% and 41.7-90.4\% without processing and floors of 3.4-3.5\% and 4.1-4.2\%, and the speaker EER on VCTK rises to 0.438-0.453, compared with 0.054-0.205. Common Voice speech at every SNR, and two-talker speech from 0~dB, leave more residual speech than VCTK: Silero detects 9.4-15.8\% and 8.0-20.1\% of their speech frames after removal, compared with 8.5-9.5\% for VCTK and a floor of 7.9-8.1\% on the clean background, and both rates increase with the SNR. With two talkers, Whisper recognises 3.4-5.3\% of the reference words after removal, against a ceiling of 53.9\%, and the speaker EER is 0.402-0.438, compared with 0.226-0.320 without processing. As on ESC-50, VAD-gated muting leaks more words than the proposed model at $-5$ and 0~dB on all three sets ($p < 0.002$), but fewer at 15~dB on Common Voice (7.0\% compared with 12.0\%, $p = 0.021$) and with two talkers (3.5\% compared with 5.3\%, $p = 0.008$), where its macro-F1 is 0.112 and 0.039, compared with 0.629 and 0.275 for the proposed model.

On every unseen corpus, the Griffin-Lim control stays close to the unprocessed mixture in activity recognition, speech detection and word hit rate (Table~\ref{tab:ood}), as on LibriSpeech.

Taken together, the model trained on read LibriSpeech speech generalises to read speech from another corpus and to crowd-sourced speech recorded on participants' own devices, whereas Common Voice speech and two simultaneous talkers leave more speech that the VADs can detect, increasingly so as the speech gets louder.

\subsection{Real-world recordings (AudioHive)}
\label{sec:res-audiohive}

Table~\ref{tab:audiohive} reports the results on the real recordings. The activity classifier is a weak probe on these data. On phase 1, recorded with the instruction not to speak, under leave-one-participant-out cross-validation, it reaches an accuracy of 0.233 and a macro-F1 of 0.177, no better than random guessing among four activities, so it does not generalise across the phase-1 homes. Trained on all phase-1 windows and applied to phase 2, it reaches an accuracy of 0.507 and a macro-F1 of 0.377 on the unprocessed audio, in which washing dishes accounts for 746 of the 1,475 windows. Small changes in activity information can therefore not be resolved on AudioHive.

Within phase 2, speech removal does not change activity recognition significantly: accuracy is 0.514 after removal, compared with 0.507 for the unprocessed audio (McNemar, Holm-corrected $p = 0.67$), and does not differ significantly from that of VAD-gated muting (0.503, $p = 0.47$) or the Facebook Denoiser (0.519, $p = 0.67$). The Griffin-Lim control lowers accuracy (0.443, $p < 10^{-7}$), whereas SepFormer raises it (0.575, $p < 10^{-7}$), although it leaves the speech proportion estimated by Silero almost unchanged.

Without a ground truth, the VADs estimate that 8.5\% (Silero), 12.9\% (WebRTC) and 20.0\% (CRDNN) of the unprocessed phase-2 audio contains speech. Speech removal lowers these estimates to 2.7\%, 4.5\% and 10.6\% (Wilcoxon, Holm-corrected $p < 10^{-60}$ for each VAD). The Facebook Denoiser reaches a similar Silero estimate (2.4\%, $p = 0.36$) but leaves more speech according to the CRDNN VAD (15.6\%) and WebRTC (5.0\%; $p < 10^{-21}$ for both). VAD-gated muting brings the estimate of Silero, which drives it, to 0.05\%, while the CRDNN VAD still estimates 13.4\% and WebRTC 19.7\%, the latter more than in the unprocessed audio.

Whisper is not a reliable measure of spoken content on these recordings. It outputs words for 98.2\% of the unprocessed phase-2 windows, although the VADs estimate that only 8.5-20.0\% of the audio contains speech, which suggests that many of its transcripts are hallucinations: in the 138 unprocessed windows in which none of the three VADs detects any speech, Whisper still outputs words for 94.9\% (2.1 words on average). Speech removal lowers the number of transcribed words from 9,155 to 7,498 (Wilcoxon on the words per window, $p < 10^{-24}$) and VAD-gated muting to 6,630, but Whisper still outputs words for 98.4\% of the windows after removal. For the same reason, the share of words of the unprocessed transcript that are still recognised after removal (49.1\%, compared with 50.0\% for VAD-gated muting and 41.1\% for the Facebook Denoiser) does not measure recovered speech, and we do not use it as one. On the real recordings, the VAD estimates are therefore our main privacy evidence.

\section{Discussion}
\label{sec:discussion}

\textbf{Removing speech within the signal preserves more than removing segments.} VAD-gated muting, the solution used in our deployment, mutes the background whenever speech is present. On ESC-50 it leaks a similar share of spoken words as the proposed model at 0-15~dB and more at $-5$~dB, but it preserves less activity information at every SNR (0.306 compared with 0.479 macro-F1 at 15~dB, and 0.425 compared with 0.636 at $-5$~dB). On the ten-second SINS clips, which contain speech for 52.1\% of their duration on average, the difference is larger: the macro-F1 of VAD-gated muting is below that of the unprocessed mixture at every SNR (0.048 compared with 0.661 for the proposed model at 5~dB), and it also leaks more words at $-5$ to 5~dB. Suppressing speech within each time-frequency region keeps more of the background sounds that overlap with speech, at a similar or better level of content privacy.

\textbf{A gap to the clean background remains when speech is loud.} With speech 15~dB above the background, the proposed model reaches 0.479 macro-F1 compared with 0.685 for the clean background, while at $-5$~dB the gap is 0.049. The losses concentrate on impulsive and broadband sounds and on non-speech vocal sounds, such as a door knock, wind or a cat, which overlap with speech in time and frequency. On SINS, macro-F1 is 0.735 compared with 0.798 for the clean background at $-5$~dB and 0.520 at 15~dB, with the largest losses for dishwashing, eating and cooking.

\textbf{The loss sets the balance between privacy and utility.} The L1-only variant had the best validation SI-SDR and the strongest privacy, but preserved the least activity information at high SNRs. Signal fidelity alone is therefore not a good criterion for choosing a speech-removal model: in deployment, the choice should weigh residual speech against the activities that must still be recognised.

\textbf{Privacy beyond speech detection.} Silero VAD, Whisper and the speaker attack order the methods largely consistently. The exception is VAD-gated muting, which Silero, the detector that drives it, ranks first at every SNR, while Whisper finds more words in its output than in that of the proposed model at low SNRs. The Griffin-Lim control shows that the low detection rates after speech removal are not an artefact of reconstruction: audio that is only reconstructed remains as detectable and as intelligible as the mixture. A VAD alone would nevertheless overstate privacy. After removal, Silero detects 1.4-2.5\% of the speech frames, yet Whisper still recovers 4.7-6.6\% of the words, two to three times its chance floor, and the speaker EER of 0.41-0.45 shows that some identity information remains. On SINS, where two activities contain speech in the background, Silero's detection rate after removal does not differ from that of the clean background at $-5$ to 5~dB, so the VAD cannot separate residual speech from background speech there, whereas Whisper still recovers 5.0-7.6\% of the words, compared with a floor of 2.9-3.0\%, and the speaker EER is 0.40-0.42. These residuals bound what the method can claim: it removes most, but not all, of the spoken content and speaker identity.

\textbf{Activities defined by speech.} A speech-removal front end cannot fully preserve activities whose sound is speech. On SINS at 5~dB, social activity reaches an F1 of 0.68 after removal, compared with 0.88 on the clean background, whereas watching TV, which also contains speech, is preserved (0.91 compared with 0.93). If social interaction is itself of interest, a deployment could keep the VAD-estimated speech proportion of each window, which shows that people talk but not what they say.

\textbf{Speaker gender.} On ESC-50, speech detection, word hit rate and activity accuracy differ little between female and male speakers: after removal, Silero detects 1.3-2.5\% and 1.5-2.6\% of their speech frames and Whisper recovers 5.2-7.0\% and 4.1-6.2\% of their words, respectively, while activity accuracy is 0.46-0.62 with female and 0.47-0.65 with male speakers. The speaker EER, however, is lower for male speakers (0.382-0.438 compared with 0.432-0.467). On SINS, male speakers leave more residual speech: Whisper recovers 6.1-10.0\% of their words after removal, compared with 3.8-6.2\% for female speakers, and the speaker EER is 0.357-0.389, compared with 0.442-0.469, while activity accuracy is similar (0.516-0.734 and 0.530-0.727). We did not test these differences for significance, but they suggest that more identity information remains in male voices, which should be examined with more speakers.

\subsection{Limitations and threats to validity}
\label{sec:limitations}

Several limitations should be acknowledged. First, the main evaluation uses synthetic mixtures. Measured RIRs, calibrated SNRs and pauses between utterances bring them closer to real recordings, and speech from unseen corpora tests generalisation beyond LibriSpeech, but the only real speech comes from AudioHive, which covers four activities and nine and seven participants in its two phases. Second, privacy is measured with off-the-shelf models, under the threat model of Section~\ref{sec:threatmodel}. Third, VAD-gated muting is driven by Silero VAD, which is also one of our three detectors; we therefore also judge it by the CRDNN VAD, Whisper and the speaker attack. Fourth, AudioHive was recorded with smartphones placed on the table nearest to each activity rather than with microphones mounted on a ceiling or wall, as an installed system would use, so the distances to the activity and to the speaker were not controlled. Fifth, Whisper and the speaker attack are evaluated on a subset of 400 clips per dataset, and the speaker attack cannot be run on Common Voice, which has no speaker labels. Sixth, on AudioHive the activity classifier is no better than random guessing under leave-one-participant-out cross-validation on phase 1, so it can resolve only large changes in activity information, and Whisper's hallucinations on the real recordings leave the VAD estimates as the only privacy measure we can interpret there. Seventh, SINS comes from a single home with one occupant, and two of its activities contain speech in the background, which limits what a VAD can tell about residual speech there.

\section{Conclusions}
\label{sec:conclusions}

This paper addressed speech privacy in acoustic sensing for assisted living. We presented a speech-removal front end, a U-Net that masks the STFT of ambient audio and returns a waveform, trained entirely on synthetic mixtures of public sounds and speech, and compared it with VAD-gated muting, speech enhancement and separation models and a reconstruction-only control under group-disjoint cross-validation. On ESC-50 and SINS, the model preserved the most activity information of all compared methods at every SNR, while Whisper recovered only a few per cent of the spoken words after removal, two to three times its chance floor, and the speaker-verification EER rose towards, but did not reach, chance. Reconstruction alone hid neither speech nor its content, and VAD-gated muting lost more activity information and leaked more words when speech was quiet. The gains held for speech from corpora not seen in training, although Common Voice speech, two simultaneous talkers and louder speech left more detectable residual speech; on real home recordings, speech removal lowered the VAD-estimated speech proportion without a significant change in activity recognition. The method therefore reduces, but does not eliminate, spoken content and speaker identity in ambient audio, and its residuals should be weighed against the activities that a deployment must recognise.

Future work will evaluate the approach against adaptive adversaries that fine-tune recognition models on processed audio, extend the privacy evaluation to other inferences from residual voice, and deploy the front end on edge hardware in care homes with installed microphones and naturally occurring conversation. Privacy-preserving acoustic sensing could increase the acceptance of ambient monitoring in elderly care, a domain where privacy concerns have impeded deployment.

\section*{Ethics and informed consent}
The AudioHive: Passive Activity Sensing Data Collection for Assisted Living study, involving human participants, was reviewed and approved by the Central Research Ethics Advisory Group (CREAG) at the University of Kent. All participants provided written informed consent prior to taking part in the study and were informed of their right to withdraw, as well as how their audio data would be stored, processed and anonymised. The third-party SINS dataset \cite{dekkers2017sins} was collected by its original authors under their own ethical approvals and was used in this work in accordance with its terms of access.

\section*{Funding and acknowledgements}
This research was informed by experience gained through the ADAPTIVE project (AI-based Assistive and Passive Technology for non-Invasive Elderly care), funded by Innovate UK, part of UK Research and Innovation (project reference 68239) \cite{ukri}, a collaboration between the University of Kent, MiiCare, East Kent Hospitals University NHS Foundation Trust and Bristol City Council. The authors thank the project partners and the study participants.

\section*{Declaration of competing interests}
The authors declare that they have no known competing financial interests or personal relationships that could have appeared to influence the work reported in this paper. The authors note that the research was informed by experience gained through the ADAPTIVE project, which involved collaboration with external project partners, including MiiCare; however, no competing financial interest or personal relationship influenced the design, execution, analysis or reporting of the work presented in this manuscript.

\section*{Data availability}
ESC-50 \cite{piczak2015esc}, LibriSpeech \cite{panayotov2015librispeech}, VCTK \cite{yamagishi2019vctk}, Common Voice \cite{ardila2020commonvoice} and the RIRS\_NOISES collection \cite{ko2017rir} are publicly available from their original sources. The SINS recordings we use \cite{dekkers2017sins} are publicly available as the development set of DCASE 2018 Task 5 \cite{dekkers2018dcase}, under the terms of its licence. The AudioHive recordings were collected in participants' homes and cannot be shared, because they may contain privacy-sensitive audio; this restriction is consistent with the ethics approval and informed-consent conditions under which they were collected. The code that produces every result in this paper, together with a page of audio examples, will be made publicly available on GitHub upon completion of the review process.

\section*{Declaration of generative AI and AI-assisted technologies in the writing process}
During the preparation of this work the authors used Claude (Anthropic), through the Claude Code tool, in order to implement and run the experimental pipeline, check references and draft and edit parts of the manuscript text. After using this tool, the authors reviewed and edited the content as needed and take full responsibility for the content of the published article.

\section*{CRediT authorship contribution statement}
\textbf{Pavlos Nicolaou:} Conceptualization, Methodology, Software, Validation, Formal analysis, Investigation, Data curation, Visualization, Writing: original draft, Writing: review \& editing.

\textbf{Christos Efstratiou:} Conceptualization, Methodology, Resources, Writing: review \& editing, Supervision, Project administration, Funding acquisition.

\clearpage
\begin{table}[p]
\centering
\caption{Activity recognition: macro-F1, pooled over the test folds (ESC-50: five folds, 50 classes, 2,000 clips per SNR; SINS: four folds, nine classes, 2,643 clips per SNR).}
\label{tab:utility}
\small
\begin{tabular}{@{}lccccc@{}}
\toprule
 & \multicolumn{5}{c}{SNR (dB)} \\
\cmidrule(l){2-6}
Method & $-5$ & 0 & 5 & 10 & 15 \\
\midrule
\multicolumn{6}{@{}l}{\textit{ESC-50}} \\
Clean background (ceiling) & 0.685 & 0.685 & 0.685 & 0.685 & 0.685 \\
Proposed (L1 + MR-STFT) & \textbf{0.636} & \textbf{0.593} & \textbf{0.574} & \textbf{0.526} & \textbf{0.479} \\
\quad MR-STFT loss only & 0.628 & 0.593 & 0.553 & 0.514 & 0.461 \\
\quad L1 loss only & 0.614 & 0.572 & 0.518 & 0.451 & 0.385 \\
Facebook Denoiser & 0.542 & 0.494 & 0.430 & 0.364 & 0.288 \\
VAD-gated muting & 0.425 & 0.340 & 0.324 & 0.309 & 0.306 \\
Mel + Griffin-Lim control & 0.409 & 0.328 & 0.238 & 0.160 & 0.092 \\
No processing & 0.401 & 0.317 & 0.233 & 0.157 & 0.099 \\
SepFormer & 0.380 & 0.343 & 0.299 & 0.219 & 0.158 \\
\midrule
\multicolumn{6}{@{}l}{\textit{SINS}} \\
Clean background (ceiling) & 0.798 & 0.798 & 0.798 & 0.798 & 0.798 \\
Proposed (L1 + MR-STFT) & \textbf{0.735} & \textbf{0.704} & \textbf{0.661} & \textbf{0.612} & \textbf{0.520} \\
Facebook Denoiser & 0.601 & 0.503 & 0.379 & 0.252 & 0.170 \\
VAD-gated muting & 0.301 & 0.111 & 0.048 & 0.040 & 0.040 \\
Mel + Griffin-Lim control & 0.443 & 0.308 & 0.164 & 0.080 & 0.055 \\
No processing & 0.451 & 0.320 & 0.178 & 0.089 & 0.057 \\
SepFormer & 0.291 & 0.308 & 0.233 & 0.135 & 0.079 \\
\bottomrule
\end{tabular}
\end{table}

\begin{table}[p]
\centering
\caption{Privacy on ESC-50 and SINS at three SNRs: share of ground-truth speech frames detected by Silero VAD, share of reference words recognised by Whisper (word hit rate) and EER of the speaker-verification attack; lower detection and hit rates and an EER closer to 0.5 mean more privacy. 400 clips per SNR for Whisper and the speaker attack, 2,000 (ESC-50) and 2,643 (SINS) for the VAD. On SINS, two activities contain speech in the background, which raises the floor of the VAD.}
\label{tab:privacy}
\footnotesize
\setlength{\tabcolsep}{3.3pt}
\begin{tabular}{@{}lccccccccc@{}}
\toprule
 & \multicolumn{3}{c}{Silero detection} & \multicolumn{3}{c}{Word hit rate} & \multicolumn{3}{c}{Speaker EER} \\
\cmidrule(lr){2-4}\cmidrule(lr){5-7}\cmidrule(l){8-10}
Method & $-5$ & 5 & 15 & $-5$ & 5 & 15 & $-5$ & 5 & 15 \\
\midrule
\multicolumn{10}{@{}l}{\textit{ESC-50}} \\
Speech track only (ceiling) & \multicolumn{3}{c}{0.960} & \multicolumn{3}{c}{0.963} & \multicolumn{3}{c}{0.040} \\
Clean background (floor) & 0.006 & 0.006 & 0.004 & 0.023 & 0.022 & 0.024 & 0.495 & 0.495 & 0.495 \\
\midrule
Proposed (L1 + MR-STFT) & 0.014 & 0.020 & 0.025 & 0.066 & 0.052 & 0.057 & 0.413 & 0.438 & 0.454 \\
\quad MR-STFT loss only & 0.021 & 0.035 & 0.042 & 0.078 & 0.069 & 0.074 & 0.382 & 0.430 & 0.434 \\
\quad L1 loss only & 0.009 & 0.011 & 0.014 & 0.038 & 0.031 & 0.035 & 0.450 & 0.468 & 0.478 \\
Facebook Denoiser & 0.107 & 0.243 & 0.474 & 0.211 & 0.224 & 0.287 & 0.286 & 0.255 & 0.236 \\
VAD-gated muting & 0.005 & 0.002 & 0.001 & 0.139 & 0.045 & 0.033 & 0.425 & 0.496 & 0.483 \\
Mel + Griffin-Lim control & 0.543 & 0.881 & 0.937 & 0.693 & 0.888 & 0.950 & 0.120 & 0.060 & 0.041 \\
No processing & 0.583 & 0.896 & 0.944 & 0.693 & 0.890 & 0.951 & 0.117 & 0.055 & 0.042 \\
SepFormer & 0.420 & 0.690 & 0.831 & 0.422 & 0.560 & 0.705 & 0.235 & 0.192 & 0.166 \\
\midrule
\multicolumn{10}{@{}l}{\textit{SINS}} \\
Speech track only (ceiling) & \multicolumn{3}{c}{0.964} & \multicolumn{3}{c}{0.973} & \multicolumn{3}{c}{0.009} \\
Clean background (floor) & 0.081 & 0.081 & 0.080 & 0.030 & 0.030 & 0.029 & 0.498 & 0.498 & 0.498 \\
\midrule
Proposed (L1 + MR-STFT) & 0.077 & 0.085 & 0.083 & 0.050 & 0.068 & 0.076 & 0.404 & 0.409 & 0.413 \\
Facebook Denoiser & 0.084 & 0.186 & 0.422 & 0.091 & 0.178 & 0.336 & 0.357 & 0.261 & 0.204 \\
VAD-gated muting & 0.002 & 0.001 & 0.002 & 0.223 & 0.091 & 0.064 & 0.377 & 0.480 & 0.487 \\
Mel + Griffin-Lim control & 0.379 & 0.799 & 0.924 & 0.561 & 0.886 & 0.956 & 0.157 & 0.049 & 0.012 \\
No processing & 0.431 & 0.842 & 0.939 & 0.559 & 0.892 & 0.958 & 0.152 & 0.040 & 0.012 \\
SepFormer & 0.318 & 0.579 & 0.778 & 0.260 & 0.502 & 0.707 & 0.240 & 0.165 & 0.125 \\
\bottomrule
\end{tabular}
\end{table}

\begin{table}[p]
\centering
\caption{Speech from unseen corpora, with ESC-50 and SINS backgrounds at 5~dB and the models trained on LibriSpeech train-clean-100: macro-F1, share of ground-truth speech frames detected by Silero VAD (Silero), Whisper word hit rate and speaker-verification EER. The ceiling row is the reverberant speech track alone; for two talkers, the reference contains the words of both talkers. Common Voice has no speaker labels. Per set, 2,000 (ESC-50) and 2,643 (SINS) clips for macro-F1 and the VAD, 400 for Whisper and the speaker attack; one silent Common Voice clip is excluded from the SINS privacy measures.}
\label{tab:ood}
\scriptsize
\setlength{\tabcolsep}{4pt}
\begin{tabular}{@{}lcccccccc@{}}
\toprule
 & \multicolumn{4}{c}{ESC-50} & \multicolumn{4}{c}{SINS} \\
\cmidrule(lr){2-5}\cmidrule(l){6-9}
 & Macro- & & Hit & & Macro- & & Hit & \\
Method & F1 & Silero & rate & EER & F1 & Silero & rate & EER \\
\midrule
\multicolumn{9}{@{}l}{\textit{LibriSpeech test-clean}} \\
Speech track only (ceiling) & -- & 0.960 & 0.963 & 0.040 & -- & 0.964 & 0.973 & 0.009 \\
Clean background & 0.685 & 0.006 & 0.022 & 0.495 & 0.798 & 0.081 & 0.030 & 0.498 \\
Proposed & \textbf{0.574} & 0.020 & 0.052 & 0.438 & \textbf{0.661} & 0.085 & 0.068 & 0.409 \\
Facebook Denoiser & 0.430 & 0.243 & 0.224 & 0.255 & 0.379 & 0.186 & 0.178 & 0.261 \\
VAD-gated muting & 0.324 & 0.002 & 0.045 & 0.496 & 0.048 & 0.001 & 0.091 & 0.480 \\
Mel + Griffin-Lim control & 0.238 & 0.881 & 0.888 & 0.060 & 0.164 & 0.799 & 0.886 & 0.049 \\
No processing & 0.233 & 0.896 & 0.890 & 0.055 & 0.178 & 0.842 & 0.892 & 0.040 \\
SepFormer & 0.299 & 0.690 & 0.560 & 0.192 & 0.233 & 0.579 & 0.502 & 0.165 \\
\midrule
\multicolumn{9}{@{}l}{\textit{VCTK}} \\
Speech track only (ceiling) & -- & 0.820 & 0.968 & 0.072 & -- & 0.822 & 0.983 & 0.020 \\
Clean background & 0.685 & 0.005 & 0.021 & 0.525 & 0.798 & 0.081 & 0.035 & 0.502 \\
Proposed & \textbf{0.619} & 0.012 & 0.055 & 0.480 & \textbf{0.711} & 0.091 & 0.080 & 0.438 \\
Facebook Denoiser & 0.501 & 0.236 & 0.192 & 0.320 & 0.464 & 0.181 & 0.162 & 0.337 \\
VAD-gated muting & 0.465 & 0.002 & 0.043 & 0.491 & 0.065 & 0.002 & 0.088 & 0.488 \\
Mel + Griffin-Lim control & 0.325 & 0.775 & 0.885 & 0.115 & 0.258 & 0.726 & 0.904 & 0.082 \\
No processing & 0.302 & 0.783 & 0.896 & 0.100 & 0.270 & 0.752 & 0.905 & 0.089 \\
SepFormer & 0.353 & 0.602 & 0.546 & 0.280 & 0.313 & 0.532 & 0.488 & 0.266 \\
\midrule
\multicolumn{9}{@{}l}{\textit{Common Voice}} \\
Speech track only (ceiling) & -- & 0.947 & 0.936 & -- & -- & 0.954 & 0.952 & -- \\
Clean background & 0.685 & 0.004 & 0.030 & -- & 0.798 & 0.080 & 0.041 & -- \\
Proposed & \textbf{0.585} & 0.034 & 0.069 & -- & \textbf{0.724} & 0.124 & 0.094 & -- \\
Facebook Denoiser & 0.471 & 0.257 & 0.203 & -- & 0.556 & 0.203 & 0.139 & -- \\
VAD-gated muting & 0.392 & 0.002 & 0.061 & -- & 0.168 & 0.002 & 0.108 & -- \\
Mel + Griffin-Lim control & 0.312 & 0.853 & 0.796 & -- & 0.413 & 0.757 & 0.767 & -- \\
No processing & 0.325 & 0.870 & 0.801 & -- & 0.447 & 0.798 & 0.774 & -- \\
SepFormer & 0.306 & 0.761 & 0.579 & -- & 0.324 & 0.673 & 0.507 & -- \\
\midrule
\multicolumn{9}{@{}l}{\textit{Two talkers (LibriSpeech test-clean)}} \\
Speech track only (ceiling) & -- & 0.960 & 0.527 & 0.275 & -- & 0.970 & 0.539 & 0.210 \\
Clean background & 0.685 & 0.005 & 0.021 & 0.514 & 0.798 & 0.081 & 0.029 & 0.502 \\
Proposed & \textbf{0.490} & 0.048 & 0.045 & 0.440 & \textbf{0.537} & 0.125 & 0.036 & 0.421 \\
Facebook Denoiser & 0.279 & 0.475 & 0.138 & 0.340 & 0.198 & 0.328 & 0.108 & 0.337 \\
VAD-gated muting & 0.193 & 0.002 & 0.030 & 0.478 & 0.040 & 0.001 & 0.042 & 0.491 \\
Mel + Griffin-Lim control & 0.187 & 0.888 & 0.422 & 0.299 & 0.095 & 0.805 & 0.410 & 0.254 \\
No processing & 0.174 & 0.902 & 0.432 & 0.295 & 0.092 & 0.844 & 0.415 & 0.255 \\
SepFormer & 0.189 & 0.801 & 0.264 & 0.334 & 0.093 & 0.723 & 0.260 & 0.310 \\
\bottomrule
\end{tabular}
\end{table}

\begin{table}[p]
\centering
\caption{AudioHive phase 2 (1,475 windows of five to ten seconds from seven participants, band-limited to 4~kHz): accuracy and macro-F1 of the activity classifier trained on phase 1, the speech proportion estimated by each VAD (no ground truth exists) and the total number of words in Whisper's transcripts, which include hallucinations. On phase 1, recorded with the instruction not to speak, leave-one-participant-out accuracy is 0.233 and macro-F1 0.177.}
\label{tab:audiohive}
\footnotesize
\setlength{\tabcolsep}{4pt}
\begin{tabular}{@{}lcccccc@{}}
\toprule
 & \multicolumn{2}{c}{Activity} & \multicolumn{3}{c}{Estimated speech proportion} & Whisper \\
\cmidrule(lr){2-3}\cmidrule(lr){4-6}\cmidrule(l){7-7}
Method & Accuracy & Macro-F1 & Silero & WebRTC & CRDNN & words \\
\midrule
Proposed & 0.514 & 0.367 & 0.027 & 0.045 & 0.106 & 7,498 \\
Facebook Denoiser & 0.519 & 0.382 & 0.024 & 0.050 & 0.156 & 8,641 \\
VAD-gated muting & 0.503 & 0.340 & 0.001 & 0.197 & 0.134 & 6,630 \\
Mel + Griffin-Lim control & 0.443 & 0.349 & 0.082 & 0.070 & 0.176 & 9,225 \\
No processing & 0.507 & 0.377 & 0.085 & 0.129 & 0.200 & 9,155 \\
SepFormer & 0.575 & 0.424 & 0.081 & 0.076 & 0.239 & 8,571 \\
\bottomrule
\end{tabular}
\end{table}
\clearpage

\appendix
\setcounter{table}{0}
\renewcommand{\theHtable}{A.\arabic{table}}
\section{Per-class precision and recall}
\label{app:perclass}

Tables~\ref{tab:perclass} and \ref{tab:perclass-esc50} give the per-class precision and recall on SINS and ESC-50 with speech 5~dB above the background; the per-class results for every method and SNR are released with the code.

\begin{table}[h]
\centering
\caption{SINS per-class precision (P) and recall (R) at 5~dB, pooled over the four test folds. Without processing, many mixtures are classified as social activity, which gives that class a high recall and a low precision.}
\label{tab:perclass}
\footnotesize
\begin{tabular}{@{}lrcccccc@{}}
\toprule
 & & \multicolumn{2}{c}{Clean background} & \multicolumn{2}{c}{No processing} & \multicolumn{2}{c}{Proposed} \\
\cmidrule(lr){3-4}\cmidrule(lr){5-6}\cmidrule(l){7-8}
Activity & Clips & P & R & P & R & P & R \\
\midrule
Absence & 300 & 0.65 & 0.80 & 0.52 & 0.04 & 0.67 & 0.57 \\
Cooking & 300 & 0.92 & 0.89 & 0.81 & 0.09 & 0.95 & 0.54 \\
Dishwashing & 300 & 0.75 & 0.72 & 0.43 & 0.03 & 0.68 & 0.38 \\
Eating & 300 & 0.83 & 0.79 & 0.12 & 0.05 & 0.49 & 0.74 \\
Other & 300 & 0.59 & 0.54 & 0.27 & 0.02 & 0.43 & 0.42 \\
Social activity & 300 & 0.93 & 0.84 & 0.13 & 0.83 & 0.60 & 0.79 \\
Vacuum cleaning & 243 & 1.00 & 1.00 & 1.00 & 0.33 & 1.00 & 0.97 \\
Watching TV & 300 & 0.90 & 0.96 & 0.42 & 0.54 & 0.87 & 0.95 \\
Working & 300 & 0.65 & 0.65 & 0.43 & 0.01 & 0.52 & 0.61 \\
\midrule
Macro average & 2,643 & 0.80 & 0.80 & 0.46 & 0.22 & 0.69 & 0.66 \\
\bottomrule
\end{tabular}
\end{table}

\begin{table}[p]
\centering
\caption{ESC-50 per-class precision (P) and recall (R) at 5~dB, pooled over the five test folds (40 clips per class).}
\label{tab:perclass-esc50}
\scriptsize
\begin{tabular}{@{}lcccccc@{}}
\toprule
 & \multicolumn{2}{c}{Clean background} & \multicolumn{2}{c}{No processing} & \multicolumn{2}{c}{Proposed} \\
\cmidrule(lr){2-3}\cmidrule(lr){4-5}\cmidrule(l){6-7}
Class & P & R & P & R & P & R \\
\midrule
Airplane & 0.67 & 0.65 & 0.44 & 0.17 & 0.50 & 0.60 \\
Breathing & 0.65 & 0.55 & 0.11 & 0.15 & 0.51 & 0.55 \\
Brushing teeth & 0.84 & 0.78 & 0.85 & 0.28 & 0.70 & 0.78 \\
Can opening & 0.54 & 0.80 & 0.00 & 0.00 & 0.38 & 0.62 \\
Car horn & 0.67 & 0.72 & 0.09 & 0.03 & 0.59 & 0.42 \\
Cat & 0.47 & 0.50 & 0.11 & 0.03 & 0.21 & 0.30 \\
Chainsaw & 0.85 & 0.88 & 0.95 & 0.45 & 0.91 & 0.75 \\
Chirping birds & 0.79 & 0.78 & 0.68 & 0.33 & 0.82 & 0.78 \\
Church bells & 0.93 & 0.95 & 1.00 & 0.25 & 1.00 & 0.80 \\
Clapping & 0.89 & 0.78 & 0.44 & 0.53 & 0.91 & 0.75 \\
Clock alarm & 0.78 & 0.78 & 0.75 & 0.15 & 0.79 & 0.68 \\
Clock tick & 0.61 & 0.78 & 0.75 & 0.07 & 0.48 & 0.30 \\
Coughing & 0.43 & 0.50 & 0.04 & 0.23 & 0.39 & 0.33 \\
Cow & 0.58 & 0.62 & 0.20 & 0.25 & 0.38 & 0.65 \\
Crackling fire & 0.64 & 0.70 & 0.75 & 0.07 & 0.58 & 0.78 \\
Crickets & 0.71 & 0.60 & 1.00 & 0.03 & 0.73 & 0.40 \\
Crow & 0.64 & 0.68 & 0.14 & 0.07 & 0.62 & 0.62 \\
Crying baby & 0.64 & 0.68 & 0.15 & 0.17 & 0.49 & 0.45 \\
Dog & 0.69 & 0.68 & 0.45 & 0.23 & 0.63 & 0.60 \\
Door wood creaks & 0.60 & 0.53 & 0.42 & 0.20 & 0.35 & 0.40 \\
Door wood knock & 0.77 & 0.82 & 1.00 & 0.05 & 0.75 & 0.38 \\
Drinking sipping & 0.56 & 0.50 & 0.13 & 0.33 & 0.27 & 0.50 \\
Engine & 0.69 & 0.68 & 0.11 & 0.35 & 0.62 & 0.50 \\
Fireworks & 0.76 & 0.72 & 0.04 & 0.12 & 0.70 & 0.70 \\
Footsteps & 0.67 & 0.75 & 0.56 & 0.23 & 0.60 & 0.68 \\
Frog & 0.81 & 0.65 & 0.60 & 0.07 & 0.70 & 0.70 \\
Glass breaking & 0.55 & 0.68 & 0.00 & 0.00 & 0.46 & 0.42 \\
Hand saw & 0.81 & 0.75 & 0.76 & 0.33 & 0.84 & 0.68 \\
Helicopter & 0.41 & 0.40 & 0.60 & 0.07 & 0.46 & 0.30 \\
Hen & 0.50 & 0.38 & 0.08 & 0.03 & 0.36 & 0.25 \\
Insects & 0.80 & 0.60 & 0.35 & 0.20 & 0.53 & 0.53 \\
Keyboard typing & 0.85 & 0.82 & 0.94 & 0.38 & 0.80 & 0.70 \\
Laughing & 0.55 & 0.53 & 0.05 & 0.60 & 0.59 & 0.47 \\
Mouse click & 0.53 & 0.53 & 0.50 & 0.05 & 0.53 & 0.45 \\
Pig & 0.58 & 0.45 & 0.12 & 0.28 & 0.50 & 0.40 \\
Pouring water & 0.80 & 0.93 & 0.89 & 0.42 & 0.77 & 0.85 \\
Rain & 0.87 & 0.82 & 0.87 & 0.33 & 0.79 & 0.75 \\
Rooster & 0.66 & 0.78 & 0.50 & 0.03 & 0.48 & 0.75 \\
Sea waves & 0.88 & 0.90 & 0.90 & 0.45 & 0.84 & 0.78 \\
Sheep & 0.82 & 0.68 & 0.16 & 0.42 & 0.62 & 0.57 \\
Siren & 0.83 & 0.85 & 1.00 & 0.20 & 0.78 & 0.62 \\
Sneezing & 0.61 & 0.62 & 0.06 & 0.03 & 0.42 & 0.53 \\
Snoring & 0.62 & 0.75 & 0.10 & 0.17 & 0.34 & 0.70 \\
Thunderstorm & 0.80 & 0.82 & 0.33 & 0.03 & 0.45 & 0.72 \\
Toilet flush & 0.94 & 0.85 & 0.79 & 0.65 & 0.85 & 0.85 \\
Train & 0.85 & 0.82 & 0.76 & 0.33 & 0.81 & 0.72 \\
Vacuum cleaner & 0.81 & 0.85 & 0.79 & 0.55 & 0.82 & 0.78 \\
Washing machine & 0.49 & 0.45 & 0.25 & 0.07 & 0.40 & 0.30 \\
Water drops & 0.67 & 0.45 & 0.25 & 0.03 & 0.57 & 0.30 \\
Wind & 0.45 & 0.55 & 0.00 & 0.00 & 0.33 & 0.17 \\
\midrule
Macro average & 0.69 & 0.69 & 0.46 & 0.21 & 0.60 & 0.57 \\
\bottomrule
\end{tabular}
\end{table}

\clearpage
\bibliographystyle{elsarticle-num}
\bibliography{ms}
\end{document}